\documentclass[twocolumn,aps,floatfix,superscriptaddress]{revtex4-2}
\usepackage[utf8]{inputenc}

\usepackage[inline]{enumitem}
\usepackage{graphicx,color}
\usepackage{dcolumn}
\usepackage{epsfig}
\usepackage{mathptmx}
\usepackage{euscript}
\usepackage{amssymb}
\usepackage{amsmath}
\usepackage{amsthm}
\usepackage{newlfont}
\usepackage{bm,mathrsfs}
\usepackage{subfigure}
\usepackage{xcolor}
\usepackage{changes}
\setaddedmarkup{\textcolor{red}{#1}}
\usepackage{todonotes}
\usepackage{subfigure}
\usetikzlibrary{calc}
\usepackage[colorlinks,linkcolor=blue,urlcolor=blue,citecolor=blue]{hyperref}
\usetikzlibrary{arrows.meta}
\newcommand{\opunit}{\textrm{1}\kern-0.22em\textrm{l}}

\usepackage{makeidx}
\usepackage{comment}

\def\bea{\begin{eqnarray}}
\def\eea{\end{eqnarray}}
\def\ba{\begin{array}}
\def\ea{\end{array}}

\newcommand{\eref}[1]{Eq.~(\ref{#1})}
\newcommand{\erefs}[1]{Eqs.~(\ref{#1})}

\newcommand{\fref}[1]{Fig.~\ref{#1}}
\newcommand{\Fref}[1]{Figure~\ref{#1}}
\def\bea{\begin{eqnarray}}
\def\eea{\end{eqnarray}}
\def\ba{\begin{array}}
\def\ea{\end{array}}

\def\la{\langle}
\def\ra{\rangle}

\definecolor{dgreen}{rgb}{0,0.7,0}

\begin{document}

\title{Universal winding laws in chiral active motion}

\author{Ion Santra}
\affiliation{Institut f\"ur Theoretische Physik, Universit\"at G\"ottingen, 37077 G\"ottingen,  Germany}
\affiliation{Department of Physics and Astronomy, KU Leuven, 3001 Leuven, Belgium }
\author{Debraj Dutta}
\author{Urna Basu}
\affiliation{S. N. Bose National Centre for Basic Sciences, Kolkata 700106, India}
\author{Sanjib Sabhapandit}
\affiliation{Raman Research Institute, Bengaluru 560080, India}

\begin{abstract}

Chiral active particles, from bacteria and sperm cells to engineered microswimmers, undergo persistent motion that breaks both time-reversal and mirror symmetries. This broken mirror symmetry leaves geometric signatures in trajectories that cannot be captured by conventional measures like position fluctuations. Here we identify the signed area and winding angle as key observables to characterize the geometric and topological aspects of the trajectories of chiral active particles. Combining theory, experiment, and computer simulations, we demonstrate that these observables exhibit universal growth laws and fluctuation statistics that are independent of propulsion details.  The presence of chirality generates multiple off-centred closed loops in the trajectories, resulting in a linear temporal growth of the signed area and a much slower logarithmic growth for the winding angle. Our work uncovers universal trajectory-level signatures of chirality and provides a unifying description of chiral dynamics in active matter.
\end{abstract}

\maketitle

Chiral active particles, ranging from rotating microorganisms to synthetic microswimmers, exhibit persistent motion with intrinsic angular dynamics that generate complex, winding trajectories~\cite{liebchen2022chiral}. Such motion is ubiquitous in living systems, including certain bacteria near substrates, spermatozoa, eukaryotic flagellates, and marine zooplankton, which exhibit intrinsic self-rotation in addition to the self-propulsion~\cite{berg90,crenshaw96,schmidt08,twsu12,lauga2006swimming,diLuzio05,nosrati15,woolley03,bohmer05,riedel05,gli08,jennings01,fenchel99,mcHenry03, friedrich2008,leonardo11}. Chiral active particles can also be artificially realized by externally driving anisotropic particles~\cite{kummel2013circular,btenhegan14,mano2017optimal,campbell17,marine13}, isotropic mini-wheels~\cite{sharma2025statistical}, as well as by forming doublets of Janus colloidal beads~\cite{ebbens10}. 
The propulsion mechanisms underlying these active motions span a broad range at the microscopic level, generating diverse types of chiral trajectories, which differ strongly in their local structure~\cite{kummel2013circular,lauga2006swimming,dey2025run,lettermann2025three}.  Nevertheless, they share a common geometric feature, namely, a persistent rotational bias. Such geometric features cannot be captured by the position fluctuations, which become effectively diffusive at long times.

\begin{figure*}[t]
    \centering
\includegraphics[width=0.9\linewidth]{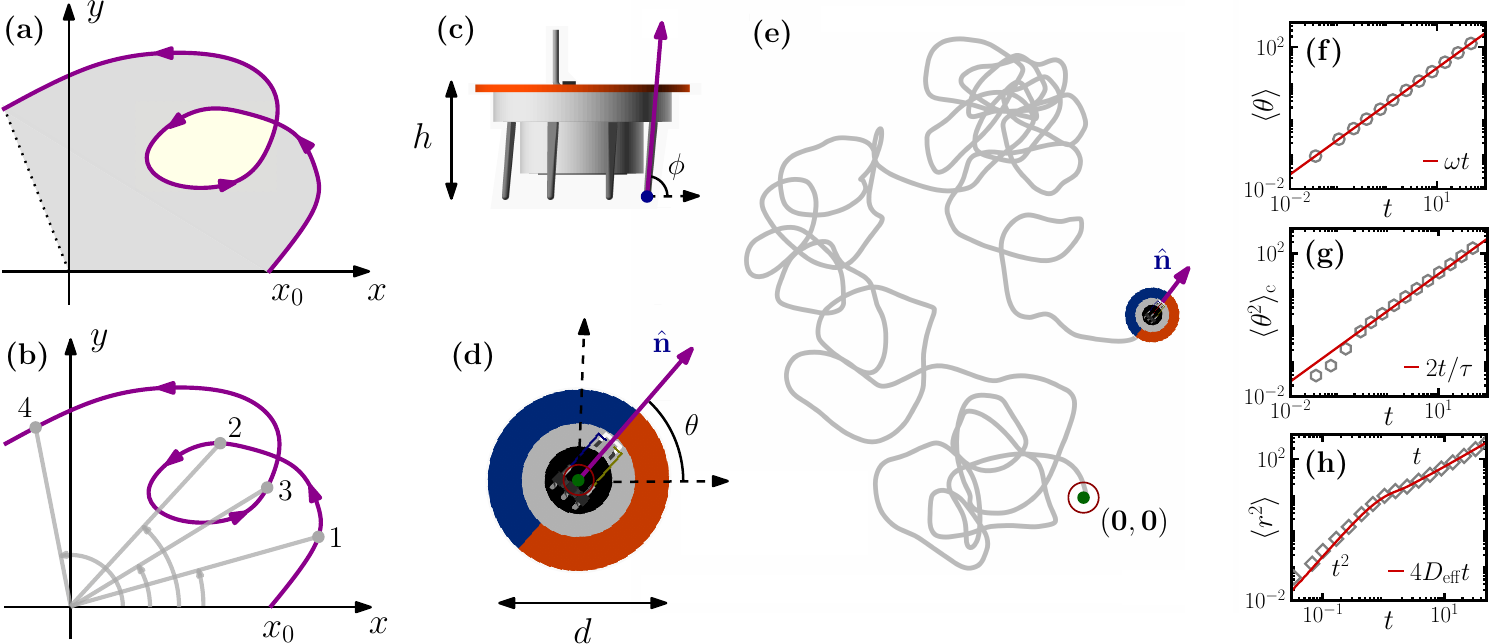}
    \caption{(a) The region shaded  gray denotes the signed area enclosed by the solid black curve on the $x-y$ plane, with respect to  the origin; the region inside the loop (shaded in yellow) contributes twice to the signed area. (b) Winding angle of the same curve about the origin with respect to the $x$-axis.  $\Omega_i$ denote the winding angle at different times along the trajectory. (c)-(d) Side view and top view of the chiral mobug with diameter $d$ and height $h$. The tilt angle of the legs is indicated by $\phi$. (e) Typical trajectory of the chiral mobug. (f) Average orientation angle $\la \theta \ra$ as a function of $t$. (g) Variance $\la \theta^2 \ra_c$ as a function of $t$. (h) Mean squared displacement $\la r^2\ra$ as a function of $t$. In all the cases symbols indicate experimental data obtained using chiral mobug. Fitting with corresponding analytic expressions (solid red lines) gives the values of the parameters $\omega, \tau$ and $D_\text{eff}$ [see \eref{eq:param}].}
    \label{fig:model}
\end{figure*}

Here we identify the signed area swept [\fref{fig:model}~(a)] and the winding angle generated by a trajectory [\fref{fig:model}~(b)] as key observables to characterize the geometric and topological aspects of chiral active motion. Combining theory, experiment, and simulations for all well-known models of chiral active motion, we show that these observables exhibit remarkable universal statistical properties. In particular, the average area grows linearly with time, whereas the average winding angle shows a logarithmic temporal growth. These growth laws are characterized by only three parameters, namely, the chirality $\omega$, the propulsion speed $v_0$, and the persistence time $\tau$. Remarkably, the probability distributions of these observables are also governed by universal scaling functions. Our results thus establish a universal geometric and topological description of chiral active motion that is independent of microscopic propulsion mechanism.

\medskip
\noindent {\it Experimental setup.}---For our experiment, we designed a self-powered chiral active particle by assembling three different components: (i) a body that holds an eccentric rotating mass (ERM) coin vibration motor, powered by a Li-Po battery attached to it, (ii) a set of tilted legs, and (iii) a `jacket' that holds the body and the legs together. The body, legs, and 
jacket are 3D printed. This motorized bug (mobug) has a diameter $d= 24$~mm and a height $h=13$~mm [Figs.~\ref{fig:model}~(c)--(d)]. 
The tilted legs of the mobug result in its active motion, with the self-propulsion speed being controlled by the tilting angle. The ERM motor has an asymmetric mass distribution, which introduces chirality to the active motion. The mobug is placed on a square glass plate (linear size 50 cm) covered with  masking tape of thickness 140 $\mu$m. 

The motion of the mobug is recorded by an overhead camera at 30 frames per second, and immediately converted to the instantaneous position (resolution  $\approx 0.5$~mm) and orientation (resolution $\approx {1.2}^{\,\circ}$)  using an object-tracking algorithm. This generates a trajectory $(x(t),y(t),\theta(t))$ of the mobug over time $t$ (\fref{fig:model}~(e); see also the  Supplemental Material (SM)~\cite{SM}). We measure the stochastic area and winding angle for each such trajectory. Their averages and distributions are then computed by sampling multiple realizations of fixed duration.

\begin{figure*}[t]
    \centering
    \includegraphics[width=0.9\linewidth]{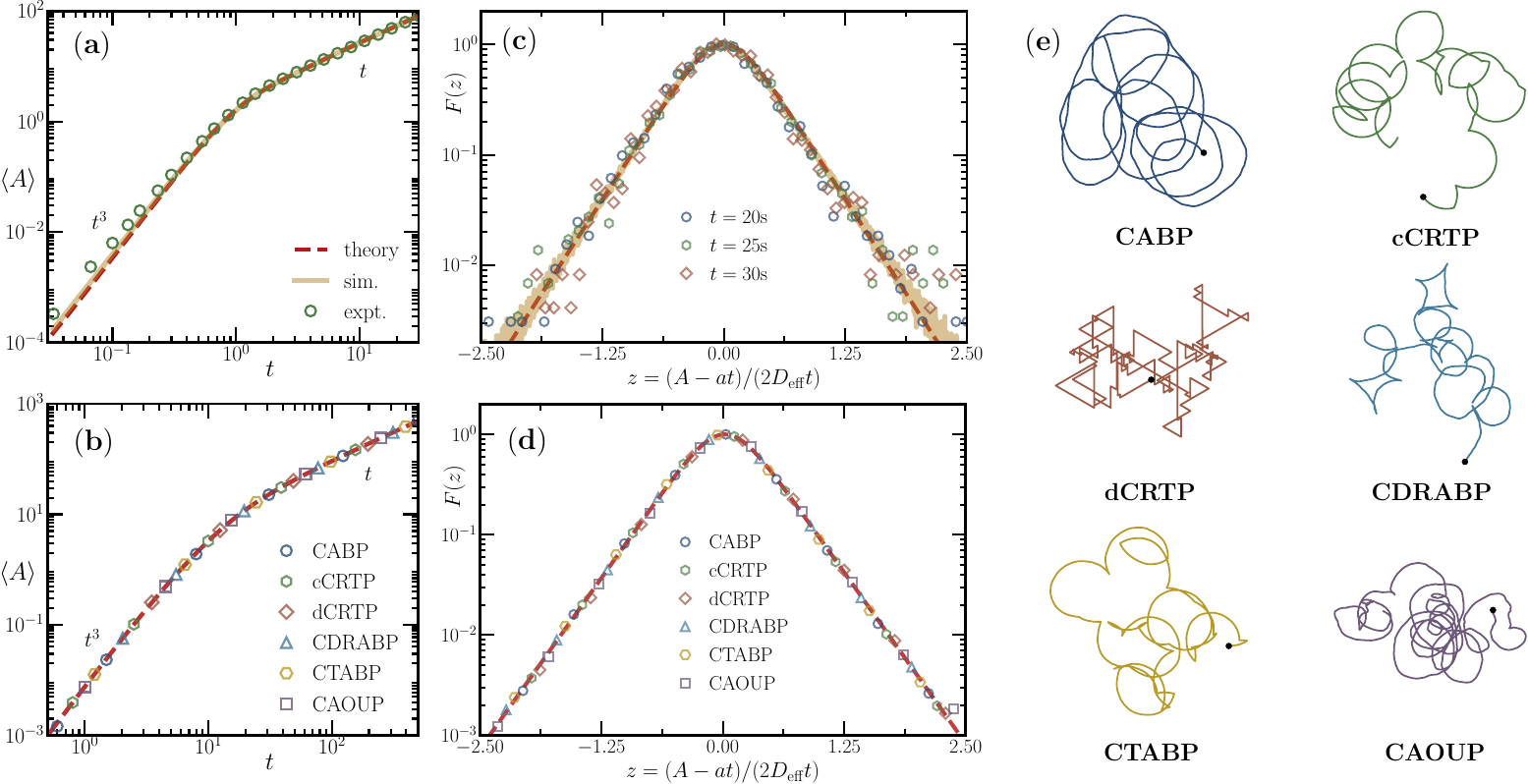}
    \caption{Behaviour of the signed area $A(t)$: (a) $\langle A(t)\rangle$ as a function of $t$. Symbols represent experimental data for the chiral mobug, while the solid and dashed lines correspond to simulation results and the analytical prediction [see \eref{eq:a:abp}], respectively evaluated with the parameter values specified in \eref{eq:param}. (b) $\langle A(t)\rangle$ versus $t$ for six different chiral active models [representative trajectories are shown in (e)], with $\tau=5.0$, $\omega=0.1$, and $v_0=1.0$. Symbols denote numerical simulation data, and the dashed line shows the analytical prediction from \eref{eq:a:abp}. (c) PDF of the scaled area $z=\frac{A-at}{2D_{\text{eff}}t}$ swept by the chiral mobug for different values of $t$,  and with $\tau, \omega$, and $v_0$ values specified in \eref{eq:param}. Solid lines indicate simulation data, while dashed lines represent the analytical result. (d) PDF of the scaled area $z$ for the different models at time $t=400$, compared with the predicted scaling function in \eref{eq:scaleA} (dashed line). Parameters used are $\tau=5.0$, $\omega=0.1$, and $v_0=1.0$.}
    \label{f:modela}
\end{figure*}

\medskip
\noindent{\it Theoretical model.}---
To theoretically model the chiral motion of the mobug, we first compute the moments of the position and orientation as functions of time, averaging over several experimental realizations.  We find that both the mean and the variance of the orientation angle grow linearly with time [Figs.~\ref{fig:model}~(f)--(g)], while the mean-squared displacement shows a crossover from ballistic to diffusive behavior [\fref{fig:model}~(h)]. These features are hallmarks of trajectories generated by a chiral active Brownian particle (CABP)~\cite{Teeffelen2008,ebbens10,kurzthaler17,sevilla2016diffusion,marine13,jahanshahi17,olsen2024optimal,Pattanayak_2024}.

A CABP moves with a constant speed $v_0$, along a self-propulsion direction $\hat{\bm n}(t) =(\cos\theta(t),\sin\theta(t))$, which undergoes an independent rotational Brownian motion with a constant drift $\omega$. Thus, the position ${\bm r}(t)$ evolves by the overdamped Langevin equation,
\begin{subequations} \label{eq:model}
\begin{align}
\dot{\bm{r}}(t)&=v_0\hat{\bm n}(t),\label{eq:langevin}\\ 
\dot{\theta}(t) &= \omega + \sqrt{2\tau^{-1}}
\eta(t)\label{eq:langevintheta}, 
\end{align}
\end{subequations}
where $\eta(t)$ is a symmetric Gaussian white noise with $\la\eta(t)\eta(t')\ra=\delta(t-t')$ and $\tau$ denotes the persistence time. The angular velocity 
$\omega$ characterizes the chirality.  It immediately follows from \eref{eq:langevintheta} that the mean and variance of the orientation angle are given by,
\begin{align}
 \la \theta(t)\ra = \omega t,~~ \la \theta(t)^2 \ra_c \equiv \la \theta(t)^2 \ra - \la \theta(t)\ra^2 = 2 t/\tau.  \label{eq:tht_moment} 
\end{align}
The mean-squared displacement can also be computed exactly~\cite{SM} and shows a diffusive behaviour at late times,
\begin{align}
    \la r^2(t) \ra \simeq 4 D_\text{eff}\, t ~\text{~with~} D_\text{eff} = \frac{v_0^2 \tau}{2(1+\omega^2\tau^2)}. \label{eq:r_moment}
\end{align}
Fitting \erefs{eq:tht_moment} and \eqref{eq:r_moment} with the experimental data [see \fref{fig:model}~(g)-(h)], we get 
\begin{equation}
 \omega = 2.57~\text{rad/s},~~ \tau = 0.73~\text{s}, ~\text{and}, ~~v_0 \approx 4.26~\text{cm/s}. \label{eq:param}
\end{equation}
In fact, using these parameters, we find that the exact $\la r^2(t) \ra$ computed theoretically  from \eref{eq:langevin} shows excellent agreement with the experimental data for all $t$ [\fref{fig:model}~(h)].

\begin{figure*}[t]
    \centering
\includegraphics[width=1\linewidth]{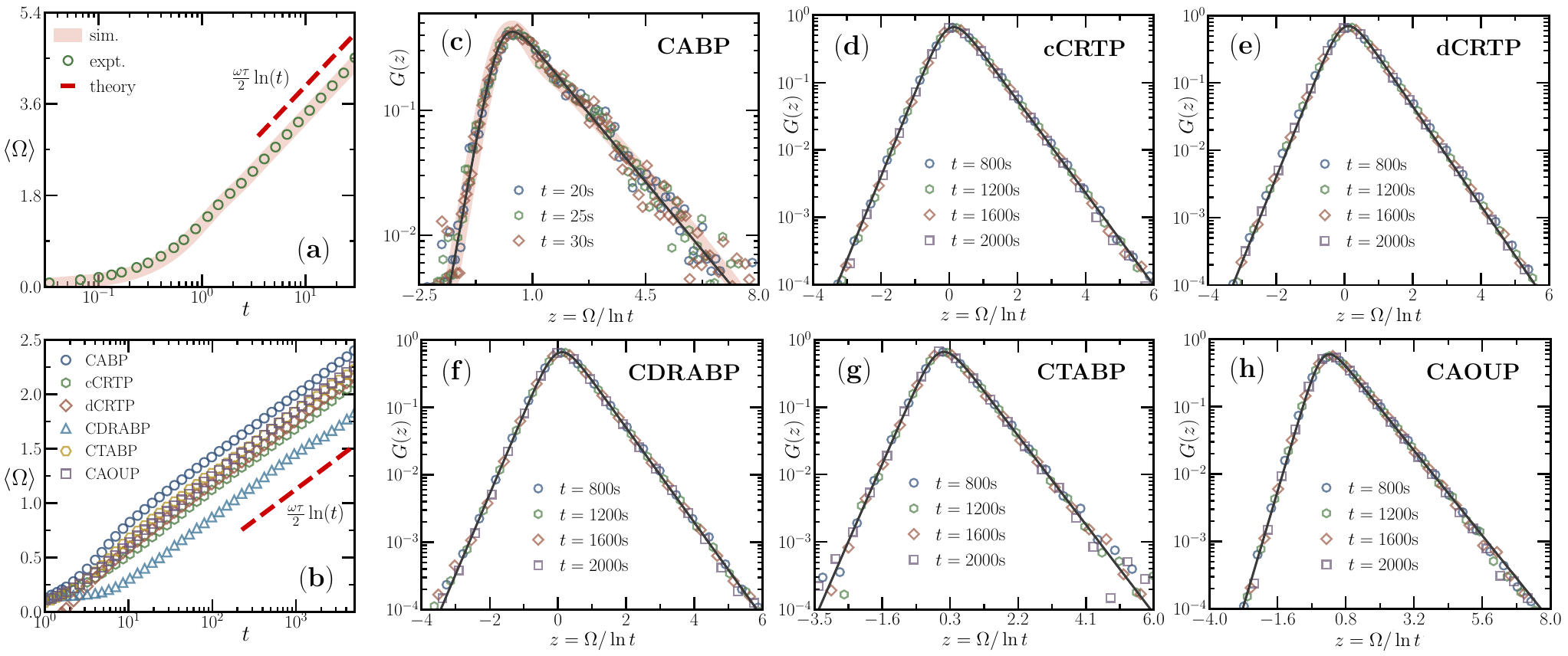}
    \caption{Behaviour of the winding angle $\Omega(t)$: (a) $\la \Omega(t) \ra$ as a function of $t$ for the CABP. Symbols represent experimental data for the chiral mobug, while the solid and dashed lines correspond to simulation results and long-time analytical prediction [see \eref{eq:winding:uni}] respectively, using parameter values given in \eref{eq:param} (b) Average winding angle for the different active particle models measured from numerical simulations with $\tau = 2, \omega = 0.25$ and $v_0 = 1$. The dashed line represents the long-time analytical prediction given in~\eref{eq:winding:uni}. (c) Distribution of the scaled winding angle for the chiral mobug: the symbols denote data for different values of $t$ and with parameter values listed in \eref{eq:param} and the solid line denotes the scaling function~\eref{gsch:def} with $\alpha=0.66, \beta=2.95$. Collapse of the scaled winding angle distribution for (d) cCRTP (e) dCRTP (f) CDRABP (g) CTABP (h) CAOUP for different values of $t$ with $\omega=1.0,\tau=0.5,v_0=1.0$. The black line in each panel corresponds to the scaling function in~\eref{gsch:def}  with $(\alpha, \beta)\simeq(1.58,2.87),\,(1.70,2.90),\,(1.62,2.75),\,(1.6,2.8)$, and $(1.21,3.28)$ respectively.}
    \label{f:modelw}
\end{figure*}

\medskip
\noindent{\it The observables.}---
The signed area swept by a planar stochastic trajectory $\{ {\bm r}(s)=(x(s),y(s)); 0 \le s \le t \}$ with respect to its initial position, is defined by the stochastic functional [see Fig.~\ref{fig:model}~(a)],
\begin{align}
    A[\{{\bm r}(s)\}] &= \frac 12 \int_0^t ds \Big[ x(s) \frac{dy}{d s} - y(s) \frac{dx}{d s} \Big]. \label{eq:def_A}
    \end{align}
   Similarly,  the winding angle about the origin with respect to the $x$-axis is defined by [see Fig.~\ref{fig:model}~(b)],
    \begin{align}
     \Omega[\{{\bm r}(s)\}] &= \int_0^t ds \frac 1 {r^2(s)}\left[ x(s) \frac{dy}{d s} - y(s) \frac{dx}{d s} \right]. \label{eq:def_Om}
\end{align}
Historically, these observables were explored through the pioneering works of L\'evy~\cite{Levy1940,Levy1950,Levy1951}, Spitzer~\cite{Spitzer1958}, and It\^{o}-McKean~\cite{Ito65} on planar Brownian motions and have been of interest in a wide range of contexts~\cite{de1971reptation,grosberg2005polymer, nelson1988vortex,duplantier1989areas}. 

From the trajectory data obtained from our experiment, we  compute the averages of area and winding angle by averaging over different realizations. We find that, at late times, the average area grows linearly [\fref{f:modela}~(a)] and the average winding angle grows logarithmically [\fref{f:modelw}~(a)] with time. We also obtain the corresponding distributions at different times [Figs.~\ref{f:modela}~(c) and \ref{f:modelw}~(c)]. We find that the distribution of the area is symmetric about the mean.  Interestingly, the winding angle follows a skewed distribution.

The calculation of the averages of the area and the winding angle depend crucially on the cross-correlation of the orientation, which can be obtained from \eref{eq:langevintheta} as~\cite{SM}, 
\begin{equation}
\la \cos \theta(t_1) \sin \theta(t_2) \ra=  - \frac{1}{2} \sin [ \omega (t_1 - t_2) ]\, e^{- |t_1 - t_2|/\tau}. \label{eq:nxny}
\end{equation}
The correlations of the orientation vector, with the parameters \eqref{eq:param}, agree with the experimental data~\cite{SM}, further validating \eref{eq:model} as an appropriate description of the mobug dynamics. The anti-symmetric nature of this cross-correlation is a direct consequence of non-zero chirality, and is responsible for the non-trivial winding properties. In particular, the chirality leads to a non-zero average area,
\begin{align}
 \la  A(t) \ra &=a\, t-b+[b\cos(\omega t)-c\sin(\omega t)]\,e^{-t/\tau},\label{eq:a:abp}
\end{align} 
where $a = \omega \tau D_\text{eff}$, $b=2D_\text{eff}\,\omega \tau^2/(1+ \omega^2\tau ^2)$, and  $c =D_\text{eff}\,\tau (\omega^2\tau^2-1)/(1+ \omega ^2\tau ^2)$. Therefore, at large times $t\gg\tau$, the average area swept increases linearly with time with a constant areal velocity $a$, while  $\la A(t)\ra\simeq (v_0^2\omega/12)\, t^3$ at short times $t\ll\tau$. We note that the areal velocity $a = \omega \tau D_\text{eff}$ is precisely the odd diffusivity that was studied recently in the context of chiral random motion~\cite{Hargus2021}. 

We compare the experimentally observed average area with our theoretical prediction in \eref{eq:a:abp} using the experimental parameters in \eref{eq:param}. We also compute the average area from numerical simulations of the stochastic dynamics \eref{eq:langevin} with these parameters. \Fref{f:modela}~(a) illustrates excellent agreement among experiment, simulation and theory.

We also analytically compute the variance of the area for the CABP model, which shows a quadratic growth $\la A^2(t)\ra_c \simeq  (D_{\text{eff}}\,t)^2$ at late times. The late-time behaviour of the mean and variance of the area suggests a scaling law for the area distribution, 
\bea
P(A,t) = \frac 1{2 D_{\text{eff}} t} F \left (\frac{A - a t}{2 D_{\text{eff}}\,t} \right). \label{eq:sc_area}
\eea
Using the data from the experiment as well as simulations, we find that the area distribution at different times shows an excellent collapse  following the scaling form \eref{eq:sc_area}. Moreover, the collapsed curve fits extremely well with 
the scaling function,
\begin{align}
    F(z) = \text{sech} (\pi z). \label{eq:scaleA}
\end{align}
This scaling function turns out to be the same as in the case of ordinary Brownian motion, although  the average area vanishes there~\cite{Levy1950}.

The other key observable carrying the signature of chirality is the winding angle $\Omega(t)$ defined in Eq.~\eqref{eq:def_Om}. Although it is hard to compute the average winding angle exactly, its large-time asymptotic behavior can be obtained as,
\begin{equation}
\la \Omega(t) \ra \simeq \frac{\omega \tau}{2}\, \ln t.
\label{eq:winding:uni}
\end{equation} 
This logarithmic growth describes the late-time behaviour of experimental data accurately, as shown in \fref{f:modelw} (a). We also measure the average winding angle from the numerical simulations with the experimental parameters in \eref{eq:param}. The excellent agreements between the experiment and simulation results at all times [see \fref{f:modelw}~(a)] confirms that the topological aspects of the trajectories of the mobugs are also accurately captured by the CABP model.

Inspired by the logarithmic time-dependence of the mean winding angle in \eref{eq:winding:uni} and the skewed nature of the distribution, we propose the scaling law
\begin{equation}
P(\Omega,t) = \frac 1{\ln t}  G\left(\frac{\Omega}{\ln t}\right), \label{eq:P_omega_sc}
\end{equation}
for the distribution of the winding angle. We find that both experimental and simulation data show excellent collapse to this scaling form [\fref{f:modelw}~(c)]. Based on extensive numerical evidence, we conjecture the scaling function as,
\begin{align}
    G(z)=\frac{1} {\pi} \sin \left(\frac{\pi \beta}{\alpha + \beta}\right)  \frac{(\alpha + \beta) }{ (e^{\alpha z} + e^{-\beta z})}~,\label{gsch:def}
\end{align}
where the $\alpha$ and $\beta$ are positive constants, and $\alpha \ne \beta$ in general. We find that $\alpha=0.66$ and $\beta=2.95$ describe both the experimental collapse and the CABP simulations performed with the experimental parameters \eqref{eq:param}.

The contrast between the linear growth of the mean signed area and a much slower logarithmic growth of the mean winding angle can be understood heuristically as follows. In the presence of a nonzero angular velocity $\omega$, typical trajectories of active particles consist of multiple closed loops that are not centered around the origin [Figs.~\ref{fig:model}~(e) and \ref{f:modela}~(e)]. Such loops contribute additively to the enclosed area [\fref{fig:model}~(a)], leading to a linear increase of the mean area with time. In contrast, a closed loop that does not encircle the origin does not contribute to the winding angle [see \fref{fig:model}~(b)]. Consequently, the winding angle accumulates only along the trajectory segments connecting successive loops that are not centered around the origin, resulting in a much slower logarithmic growth. This physical argument is quite generic as it does not rely on any specific details of the model. Therefore, a natural question is: how robust are these geometric and topological laws?

\medskip
\noindent{\it Universality.}---
Experimentally probing the question of robustness for a broad class of chiral dynamics is challenging. Instead, we take recourse to computer simulations of a wide variety of chiral active particles ranging from continuous and discrete chiral run-and-tumble particles (cCRTP and dCRTP)~\cite{mallikarjun2023chiral,angelani2022orthogonal,larralde97} to chiral active Ornstein-Uhlenbeck particle (CAOUP)~\cite{Caprini-caoup}. We also introduce the chiral versions of direction reversing active Brownian particles (CDRABP) ~\cite{santra2021active} and a
tumbling active Brownian particle (CTABP). These models are explicitly defined in Methods; also see Fig.~\ref{f:modela}~(e) for their typical trajectories.   

We find that the laws governing the winding observables \eqref{eq:def_A} and \eqref{eq:def_Om} remain valid for all these models. Despite their fundamentally different microscopic mechanisms of self-propulsion, these laws can be expressed in terms of the same three effective parameters, namely, the chirality $\omega$, the persistence time $\tau$, and the characteristic self-propulsion speed $v_0$. The universality of the linear and logarithmic growth laws for the average area [\fref{f:modela}\,(b)] and winding angle [\fref{f:modelw}\,(b)], respectively, can be understood theoretically by computing the cross-correlation of the orientation vector, which, expressed in terms of these parameters, admits a universal form  \eqref{eq:nxny} across all models. The universality also extends to the fluctuations of the area and the winding angle. Not only does the scaling law \eqref{eq:sc_area} hold for area fluctuations of all the models, but also the scaling function remains the same as in \eref{eq:scaleA} [\fref{f:modela}(d)]. The winding angle distribution also follows the universal scaling law \eqref{eq:P_omega_sc} with the conjectured scaling function in \eqref{gsch:def}, albeit with model-dependent parameters $(\alpha,\beta)$ [\fref{f:modelw}(d)-(h)]. The geometric and topological laws exhibited by the mobug are thus universal,  thereby providing a unifying framework for understanding chiral active motion.

Since an active motion typically becomes diffusive at late times, it is instructive to contrast the topological properties of chiral active motion with that of Brownian motion. 
Unlike \eref{gsch:def}, the scaling function associated with the winding angle distribution of the Brownian motion~\cite{Spitzer1958} is given by the Cauchy distribution $G(z)=2/[\pi (1+4z^2)]$. The power-law tail $G(z) \sim z^{-2}$ originates from the diverging windings of Brownian motion around the origin in a finite time. This can be cured by excluding the trajectory segments near the origin, putting an obstacle at the origin, or considering random walks with bounded jumps, which lead to exponential tails~\cite{pitman1986level, belisle1989windings, comtet1993asymptotic, comtet1993asymptotic2, drossel1996winding, wen2019winding,rudnick1987winding,drossel1996winding,saleur1994winding,grosberg2003winding}. Our work reveals that active motion naturally eliminates the pathology of infinite windings since the persistent nature of the active motion prevents the particle from remaining infinitesimally close to the origin for long. 

\medskip
\noindent{\it Conclusions and outlook.}---
In conclusion, combining experiment, theory and simulations, we have characterized chiral active motion using a geometric observable (signed area) and a topological observable (winding angle) in terms of three physical parameters $(v_0, \omega, \tau)$, and discovered a striking universality across diverse active particle models.  Since both area and winding angle are time-integrated observables, our universal laws enable an accurate estimation of the parameters from experimental data, especially for smaller particles where tracking the self-propulsion direction is  challenging~\cite{hernandez2020collective}.
We believe that our work will stimulate broader explorations of geometry and topology in active matter systems. \\

\acknowledgements

D.D. thanks S. Shahid and A. Chaudhuri for insightful discussions on experimental design.
U.B. thanks S. Das for the discussions at the initial stage of the work. I.S. acknowledges funding from the European Union’s Horizon 2024 research and innovation programme under the Marie Sklodowska-Curie (HORIZON-TMA-MSCA-PF-EF) grant agreement No. 101205210.
U.B. acknowledges the support from the Anusandhan National Research Foundation (ANRF), India, under a MATRICS grant [No. MTR/2023/000392]. This research was supported in part by grant NSF PHY-2309135 to the Kavli Institute for Theoretical Physics (KITP).

\onecolumngrid 

\bigskip \bigskip 
\centerline {\bf \large End Matter}
\bigskip 

\twocolumngrid

\appendix

\section{Experiment} 

\noindent {\bf Particle fabrication.}---
The components of the mobug are 3D printed in polylactic acid (PLA) via fused deposition modelling (Ender 3 Pro, nominal print precision 0.075 mm). The compliant legs are initially printed flat and subsequently wrapped around the body and secured by the jacket. To induce directed motion under vibration, the legs are fabricated with a controlled tilt angle of $5^\circ$. For robust optical tracking of both position and orientation, a high-contrast, bi-colored annular marker is placed on top of the jacket.
To improve surface finish and mechanical durability, the printed components are manually polished, coated with a thin layer of nitrocellulose, and subsequently UV-cured to enhance rigidity and wear resistance.  

\noindent{\bf Setup and data acquisition.}--- The mobug is released on a square glass plate coated with a layer of saturated masking paper to ensure uniform surface properties. The plate is mounted on a levelling platform, and its tilt is adjusted  to minimize any systematic drift. 

The motion of the mobug is recorded using an overhead camera. The video stream is acquired at 30 frames per second and processed in real time using a custom Python pipeline. After cropping, frames are analyzed at a resolution of $1000 \times 1000$ pixels. Using OpenCV-based image processing, the particle position and orientation are extracted from each frame, yielding trajectories of the particles. In total, approximately 90 minutes of trajectory data were collected.

For statistical analysis, the full trajectory of duration $[0,T]$ is segmented into overlapping sub-trajectories of fixed duration $t$, with a shift interval $\delta t \approx 150\,\mathrm{ms}$, which is significantly larger than the characteristic correlation time  $\approx 50\,\mathrm{ms}$ of the mobug motion.

\section{Theory}

\noindent In this section we provide details about the theoretical computation of the relevant observables. We also define different chiral active models used for computer simulations.

\subsection{Moments of area and winding angle for CABP}

\noindent For the CABP dynamics~\eqref{eq:langevin}, the average area reduces to,
\begin{align}
    \la A(t) \ra = \frac {v_0^2}2 \int_0^t d s \int_0^s d s'  \left[\la n_x(s') n_y(s)\ra - \la  n_x(s)n_y(s') \ra\right]. 
    \label{avg:area:form}
\end{align}
From \eref{eq:def_A}, we can write the second moment of the area as,
\begin{align}
     \la A^2(t)\ra &=\frac{v_0^4}{4}\int_0^t d\tau_1\int_0^t d\tau_2\int_0^{\tau_1} d\tau_3\int_0^{\tau_2} d\tau_4 \cr
     &\times \la\sin\left(\theta(\tau_1)-\theta(\tau_3)\right)\sin\left(\theta(\tau_2)-\theta(\tau_4)\right)\ra,
\end{align}
where $\theta(\tau_i)$ are correlated Gaussian random variables. Using the identity for Gaussian random variables $\{\phi_j\}$, $\la \exp[i\sum_j a_j\phi_j]\ra=\exp\left[-\frac 12\sum_{j,k}a_ja_k\la \phi_j\phi_k\ra\right]$, and performing the integrals over $\tau$ we get the second moment of the swept area exactly. The full expression is rather lengthy and not very illuminating. The leading order behavior subtracting the mean and taking the long time limit leads to $\la A^2(t)\ra_c \simeq  (D_{\text{eff}}\,t)^2$.

Although it is hard to compute the moments of the winding angle  exactly, the large-time asymptotic behaviour of the mean winding angle $\langle \Omega (t) \rangle$ can be extracted using the following approximation,
 \begin{align}
     \la \Omega(t) \ra \approx \int_0^t ds\, \frac 1{\langle r^2(s)\rangle} \langle x(s) \dot{y}(s) - y(s) \dot{x}(s)\rangle. \label{eq:Omega_approx}
 \end{align}
The numerator and denominator of the integrand can be computed exactly~\cite{SM}). For large $t$, the integral in \eref{eq:Omega_approx} is dominated by the contribution from the large $s \gg \tau$ regime where the integrand decays as $\omega \tau/ (2 s)$. Thus, at large times, the approximation \eref{eq:Omega_approx} leads to the logarithmic growth reported in \eref{eq:winding:uni}. Figure~\ref{f:modelw}\,(a) illustrates that \eref{eq:winding:uni} describes the asymptotic behaviour of the average winding angle very well.

\subsection{Models}

\noindent Apart from the chiral Active Brownian Particle  discussed in the main text, we also comprehensively study the winding properties of five other models of chiral active motion. In the following we define these models. \\

\noindent{\bf Chiral Run-and-Tumble Particles (CRTP):} The position of a Run-and-tumble particle (RTP) moving with a self-propulsion speed $v_0$ in two dimensions evolves according to the overdamped Langevin equation \eref{eq:langevin}, where $\hat{\bm{n}} = (\cos\theta, \sin\theta)$ changes intermittently via
$\theta \to \theta^\prime$ at a certain rate. Chirality is introduced through a biased dynamics of the orientation angle $\theta$. We consider two variants of chiral RTP below.  \\

\noindent {\it Continuous CRTP (cCRTP):} The orientation $\theta \in [0, 2\pi]$ evolves continuously via a deterministic rotation along with the stochastic tumblings. The tumbling events happen at a rate $\gamma$, which implies that in an infinitesimal time interval $\Delta t$, the evolution of the orientation follows,
\begin{align}
 \theta(t+\Delta t ) = 
\begin{cases}
\theta(t)+ \omega\, \Delta t & \text{with probability } (1 - \gamma \Delta t), \\
\theta' & \text{with probability } \gamma \Delta t,
\end{cases}
\label{eq:ccrtp-app}
\end{align}
where $\theta'$ is chosen uniformly from $[0, 2\pi]$. \\

\noindent{\it Discrete CRTP (dCRTP):} The orientation $\theta$ takes values from a finite set, and evolves via a Markov jump process. Chirality is introduced by using asymmetric switching rates between the states. In particular, we consider a three-state model with $\theta \in \left\{ \pi/2, 7\pi/6, 11\pi/6 \right\}$, and clockwise/anticlockwise transition rates $\gamma$ and $\lambda \gamma$ respectively, with $\lambda > 1$.


Though the chirality $\omega$ does not appear explicitly in the dynamical description, it can be identified from the correlations among the components of the orientation vector,
which have the same form as in \eqref{eq:nxny}, with
\begin{equation}
    \omega=\frac{\sqrt{3}\gamma(\lambda-1)}{2},\quad\tau=\frac{2}{3\gamma(1+\lambda)}.
\end{equation}
and $\lambda = (1+ \sqrt{3}\, \omega \tau)/(1- \sqrt{3}\, \omega \tau), \quad \gamma = (1- \sqrt{3}\, \omega \tau)/(3 \tau).$
Since $\lambda$ and $\gamma$ must be positive, there is a restriction on $\omega$ and $\tau$ for the dCRTP, namely, $\omega \tau < 1/\sqrt{3}$. \\

\noindent{\bf Chiral Direction Reversing Active Brownian Particle (CDRABP):} The direction-reversing active Brownian particle (DRABP) models a class of active dynamics where, in addition to the rotational diffusion of ABP, the particle undergoes complete directional reversals intermittently. 
The chiral version of DRABP is modeled theoretically by the Langevin equation \eref{eq:langevin} with 
$\hat{\bm{n}}(t) \equiv (n_x(t),n_y(t)) = (\sigma(t) \cos \theta(t), \sigma(t) \sin \theta(t))$, where $\sigma(t)$ is a dichotomous noise switching between $\pm 1$ at rate $\gamma$, and $\theta(t)$  evolves continuously following 
\eref{eq:langevintheta}. It can be easily checked that auto-correlation of the orientation vector $\hat{\bm{n}}(t)$ follows the same form of cross-correlations as \eqref{eq:nxny},
\begin{equation}
\la n_x (t_1) n_y(t_2) \ra=  - \frac{1}{2} \sin [ \omega (t_1 - t_2) ]\, e^{- |t_1 - t_2|/\tau},
\end{equation}
with $\tau = 1/(D_R + 2 \gamma)$. \\

\noindent{\bf Chiral Tumbling active Brownian Particle (CTABP):}
This is a generalization of the DRABP, where instead of reversing the direction, the particle can \emph{tumble} to any direction. This motion is also modeled by \eref{eq:langevin} where 
the orientation angle is divided into two independent parts, 
\begin{align}
    \theta=\theta_1+\theta_2\, ,
\end{align}
where $\theta_1$ follows the chiral ABP dynamics \eref{eq:langevintheta}, and $\theta_2$ follows 
\begin{align}
\theta_2(t+\Delta t ) = 
\begin{cases}
\theta_2(t) & \text{with probability } (1 - \gamma \Delta t), \\
\theta_2' & \text{with probability } \gamma \Delta t.
\end{cases}
\end{align}
The new $\theta_2'$ after each tumble is chosen uniformly from $[0,2\pi]$. It can be easily checked that cross-correlation of the orientation vector $\hat{\bm{n}}(t)$ follows \eqref{eq:nxny},
with $\tau = 1/(D_R +  \gamma)$. \\

\noindent{\bf Chiral Active Ornstein-Uhlenbeck Particle (CAOUP):} In contrast to the other models, the active Ornstein-Uhlenbeck particle (AOUP) features a fluctuating self-propulsion velocity ${\bm v}=(v_x, v_y)$ governed by independent Ornstein-Uhlenbeck processes. Chirality is introduced by adding a rotational coupling to the velocity dynamics~\cite{Caprini-caoup}. Thus, the position vector $\bm{r} =(x,y)$ evolves via, 
\begin{align}
\dot{x} &= v_x, \qquad \dot{y} = v_y, \nonumber \\
\tau \dot{v}_x &= -v_x - \omega \tau v_y + v_0 \sqrt{\tau}\, \xi_x(t), \nonumber \\
\tau \dot{v}_y &= -v_y + \omega \tau v_x + v_0 \sqrt{\tau}\, \xi_y(t),
\end{align}
where $\xi_x(t)$ and $\xi_y(t)$ are independent white noises with $\langle \xi_i(t) \xi_j(t') \rangle =  \delta_{ij} \delta(t - t') $. 
It is again straightforward to see that the auto-correlations of the orientation vector $\hat{\bm{n}} = \bm{v}/v_0$ are given by \eqref{eq:nxny}. \\

\section{Numerical simulations}

\noindent We use Euler discretization scheme with $dt=10^{-4}-10^{-3}$  to numerically simulate the Langevin dynamics \eref{eq:langevin} with the appropriate dynamics for $\hat {\bm n}$ for different models. Typically  $10^5-10^7$ realizations are used to measure the moments and the distributions of the relevant observables. For the statistics of area and winding angle, we use the initial positions $(x(0),y(0))=(0,0)$, and $(x(0),y(0))=(0.1,0)$, respectively.


\bibliography{2ref}

\end{document}


\title{Supplemental Material: Universal winding laws in chiral active motion}

\author{Ion Santra}
\affiliation{Institut f\"ur Theoretische Physik, Universit\"at G\"ottingen, 37077 Germany }
\affiliation{Department of Physics and Astronomy, KU Leuven, 3000 Belgium}
\author{Debraj Dutta}
\author{Urna Basu}
\affiliation{S. N. Bose National Centre for Basic Sciences, Kolkata 700106, India}
\author{Sanjib Sabhapandit}
\affiliation{Raman Research Institute, Bengaluru 560080, India}

\begin{abstract}
Here, we provide a few details related to the main text.
\end{abstract}

\maketitle

\section{Two point correlations of orientation}

\begin{figure}
    \centering
    \includegraphics[width=0.9\textwidth]{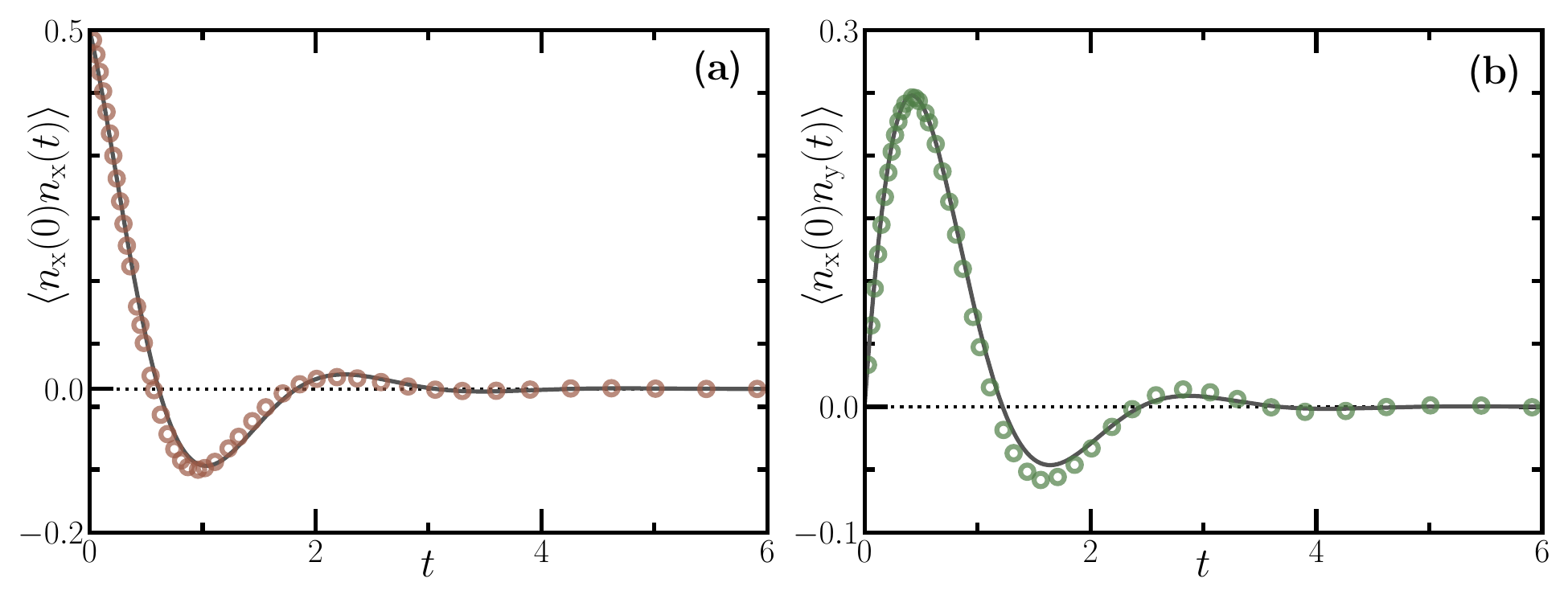}
    \caption{Two-point correlation of the internal orientation of the mobug measured in the experiment. The solid lines indicate the theoretical predictions \erefs{eq:corr:sm} and \eqref{eq:cross-corr:sm}, evaluated using the fitted parameters $\omega = 2.57~\text{rad/s}, \tau = 0.73~\text{s}$ and $v_0 \approx 4.26~\text{cm/s}$ [see Eq.~(4) in the main text] extracted from the experimental trajectories.}
    \label{fig:expt_corr}
\end{figure}

Let us recall that the dynamics of the chiral active Brownian particle in two dimensions is given by [Eq.~(1) in the main text],
\begin{equation}
\begin{split}
    \frac{dx}{dt} &= v_0 \cos\theta(t), ~ \frac{dy}{dt} = v_0 \sin\theta(t).
    \end{split}
    \label{eq:dyn}
\end{equation}
where $\theta(t)$ evolves according to,
\begin{align}
\dot{\theta}(t) &= \omega + \sqrt{2\tau^{-1}}\,
\eta(t)\label{eq:langevinthetaSM}, 
\end{align}
The Gaussian nature of the white noise ensures that the propagator for the $\theta$ process is given by,
\begin{align}
    P(\theta,t|\theta',t')=\frac{1}{\sqrt{4\pi \tau^{-1}(t-t')}}\exp\left[-\frac{(\theta-\theta'-\omega(t-t'))^2}{4\tau^{-1} (t-t')}\right].
    \label{eq:propcabp}
\end{align}
Using the above propagator, the two-point correlations of the orientation angle $\theta(t) $ can be computed in a straightforward manner. 
In the stationary state, these correlations are given by,
\begin{align}
    \langle \cos\theta(t_1) \cos\theta(t_2) \rangle =     \langle \sin\theta(t_1) \sin\theta(t_2) \rangle 
    & = \frac{1}{2} e^{- \frac{|t_1 - t_2|}{\tau}}  \cos [ \omega (t_1 - t_2) ]
    \label{eq:corr:sm} \\
    \langle \cos\theta(t_1) \sin\theta(t_2) \rangle &=  - \frac{1}{2} e^{- \frac{|t_1 - t_2|}{\tau}}  \sin [ \omega (t_1 - t_2) ], \label{eq:cross-corr:sm}
\end{align}
Figure~\ref{fig:expt_corr} illustrates that the 
experimentally observed two-point correlations of the mobug orientation show excellent  agreement with the above theoretical predictions \eref{eq:corr:sm} and \eref{eq:cross-corr:sm}. Although the two-point orientation correlations have the same form for all the chiral active particle models considered here, the underlying orientation dynamics is model dependent. This difference is seen clearly at the level of the corresponding propagators. For completeness, we list the propagators for the different models below.

For cCRTP, the propagator is,
\begin{align}
P(\theta,t|\theta',t')
=
e^{-\gamma t}
\delta(\theta-\theta'-\omega (t-t'))
+
\frac{1-e^{-\gamma (t-t')}}{2\pi}.
\label{eq:propccrtp}
\end{align}

For dCRTP, the orientation dynamics follows a Markov jump process with states $\theta_i=\left\{ \pi/2, 7\pi/6, 11\pi/6 \right\}$. The propagator is given by,
\begin{align}
P(\theta_i,t|\theta_j,t')=\left[e^{W(t-t')}\right]_{ij}, \quad \text{with~}W=
\begin{pmatrix}
-(1+\lambda)\gamma & \lambda\gamma &\gamma\\
\gamma & -(1+\lambda)\gamma &\lambda\gamma\\
\lambda\gamma& \gamma &-(1+\lambda)\gamma
\end{pmatrix}.
\end{align}

For a CDRABP, the propagator is,
\begin{align}
    P(\theta,\sigma,t|\theta',\sigma',t')=\frac{1}{2\sqrt{4\pi D_R(t-t')}}\exp\left[-\frac{(\theta-\theta'-\omega(t-t'))^2}{4D_R (t-t')}\right]\left(1+\sigma \sigma' e^{-2\gamma (t-t')}\right)
\end{align}

For CTABP, the two stochastic components of the orientation angle $\theta=\theta_1+\theta_2$ are independent, and the corresponding propagator is just a product of the individual propagators,
\begin{align}
    P(\theta,t|\theta',t')=P_1(\theta_1,t|\theta_1',t')P_2(\theta_2,t|\theta_2',t')
\end{align}
where $P_1$ and $P_2$ follow \eref{eq:propcabp} and \eref{eq:propccrtp}, respectively.

For the CAOUP model, the orientation vector is related to the propulsion velocity vector as  $\hat{\bm n}=\bm v/v_0$. Since $\bm{v}$ is a Gaussian, the propagator of the propulsion velocity vector follows as,
\begin{align}
   P(\hat{\bm n},t|\hat{\bm n}',t')=\frac{1}{2\pi \Sigma(t-t')}
\exp\left[
-\frac{
\left|
\bm {\hat n}-M(t-t')\bm {\hat n}'
\right|^2
}{
2\Sigma(t-t')
}
\right],
\end{align}
where 
\begin{align}
    M(t)=e^{-t/\tau}\begin{pmatrix}
        \cos(\omega t)& -\sin(\omega t)\\
        \sin(\omega t)& \cos(\omega t)
    \end{pmatrix}, \quad\text{and  } \Sigma(t)=\frac12\left( 1-e^{-2t/\tau}\right).
\end{align}

\section{Mean-squared Displacement}

The second moment of the position is given by [see Eq.~(1a) in the main text], 
\begin{align}
   \la r^2(s) \ra = r_0^2 + v_0^2 \int_0^s dt_1 \int_0^s dt_2 \, \la \hat{\bm n}(t_1) \cdot \hat{\bm n}(t_2) \ra   
\end{align}
where $r_0^2=x(0)^2+y(0)^2$. Substituting  Eqs.~\eqref{eq:dyn} and \eqref{eq:corr:sm}, in the above equation,  the integrals over time can be performed easily. This gives,
\begin{align}
    \langle r^2(s) \rangle &=  4 D_\text{eff} \, (s  +  \tau_0)  -\frac{ 4 D_\text{eff} \,\tau  e^{-{s}/{\tau }}}{\left(1+\omega^2\tau^2 \right)}\left[\left(\omega^2 \tau^2 -1\right) \cos (\omega s)+2 \omega \tau   \sin (\omega s)\right],~~
    \label{eq:r2:sm}
 \end{align}
where, 
\begin{align}
D_\text{eff}  = \frac{v_0^2 \tau}{2(1+\omega^2\tau^2)}\quad \text{and} \quad
    \tau_0 = \frac{(\omega^2 \tau^2 -1) \tau }{1+ \omega^2 \tau^2} +  \frac{r_0^2}{4 D_\text{eff}}\, .
    \label{eq:tau0}
\end{align}

\section{Approximate calculation of the mean winding angle}

The winding angle swept by a trajectory $\{{\bm r}(s); 0 \le s \le t\}$ is given by Eq.~(2) of the main text, which reads,
 \begin{equation}
     \Omega[{\bm r}(s)] = \int_0^t ds \frac{1} {r^2(s)}\left[ x(s) \frac{dy}{d s} - y(s) \frac{dx}{d s} \right]. \label{eq:def_Om:sm}
\end{equation}
As we have mentioned in the main text, unlike the average area, which can be computed exactly and given by Eq.~(8) of the main text, the average winding angle is hard to calculate exactly, due to the $1/r^2(s)$ factor in the integrand in Eq.~\eqref{eq:def_Om:sm}. Therefore, to proceed further, we make an approximation, where we replace the average of the integrand in  Eq.~\eqref{eq:def_Om:sm} with the ratio of the averages of the numerator and denominator, i.e., 
\begin{equation}
     \langle \Omega (t) \rangle \approx \int_0^t I(s)\, ds ,
     \label{eq:av_Om:sm}
\end{equation}
where the integrand $I(s)$ is given by,
\begin{equation}
    I(s)= \frac{1} {\langle r^2(s)\rangle}\, 
     \left\langle \left[ x(s) \frac{dy}{d s} - y(s) \frac{dx}{d s} \right]\right\rangle.
     \label{eq:integrand}
\end{equation}


Using the Langevin dynamics \eqref{eq:dyn} and the cross correlation [Eq.~(15) in the main text], we compute the second average in Eq.~\eqref{eq:integrand} exactly, as 
\begin{align}
  \left\langle \left[ x(s) \frac{dy}{d s} - y(s) \frac{dx}{d s} \right]\right\rangle =  v_0^2 \int_0^s e^{-s'/\tau} \, \sin(\omega s')\, ds'
  = 2 D_\text{eff} \,\omega\, \tau 
  - 2 D_\text{eff} \,  e^{-{s}/{\tau}}\, \bigl[  \omega \tau \cos (\omega s)+\sin (\omega s)\bigr]\, .
  \label{eq:2ad:sm}
\end{align}

\begin{figure}[t]
    \centering
    \includegraphics[width=0.49\linewidth]{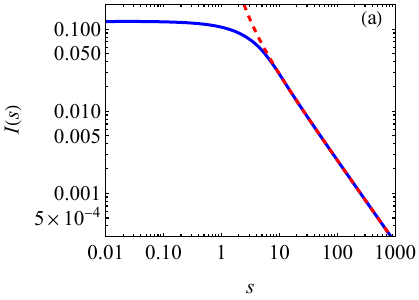} \includegraphics[width=0.47\linewidth]{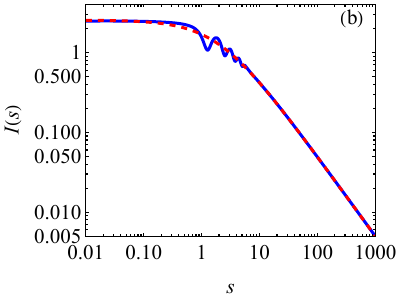}
    \caption{Behavior of the integrand $I(s)$: The blue solid lines plot the exact $I(s)$ for (a) $\omega=0.25$ and (b) $\omega=5$ with $\tau=2$, $v_0=1$, and $r_0=0.001$ for both figures. The red dashed lines plot the long time approximation given in Eq.~\eqref{eq:int-long}. }
    \label{fig:integrand}
\end{figure}

Therefore, the integrand $I(s)$ in Eq.~\eqref{eq:integrand} is given by the ratio of the expressions in Eqs.~\eqref{eq:2ad:sm} and \eqref{eq:r2:sm} respectively.  For large $t$, the dominant contribution to the integral in \eref{eq:av_Om:sm} comes from large $s \gg \tau$ regime [see \fref{fig:integrand}], where the integrand can be approximated as,
\begin{equation}
    I(s) \simeq \frac{\omega \tau}{2 (s+\tau_0)}\, 
    \label{eq:int-long}
\end{equation}
where $\tau_0$ is given in \eref{eq:tau0}.
This late-time $s^{-1}$ power-law tail of the integrand $I(s)$ immediately implies a logarithmic behavior
\begin{equation}
  \langle \Omega (t) \rangle \approx \frac{\omega \tau}{2}\, \ln t\, ,  
  \label{eq:logt}
\end{equation}
at large times, which is quoted in Eq.~(11) in the main text. As shown in  Fig.~3~(a) of main text, Eq.~\eqref{eq:logt} accurately describes the long-time growth of the average winding angle.

\begin{figure}[t]
    \centering
    \includegraphics[width=0.9\linewidth]{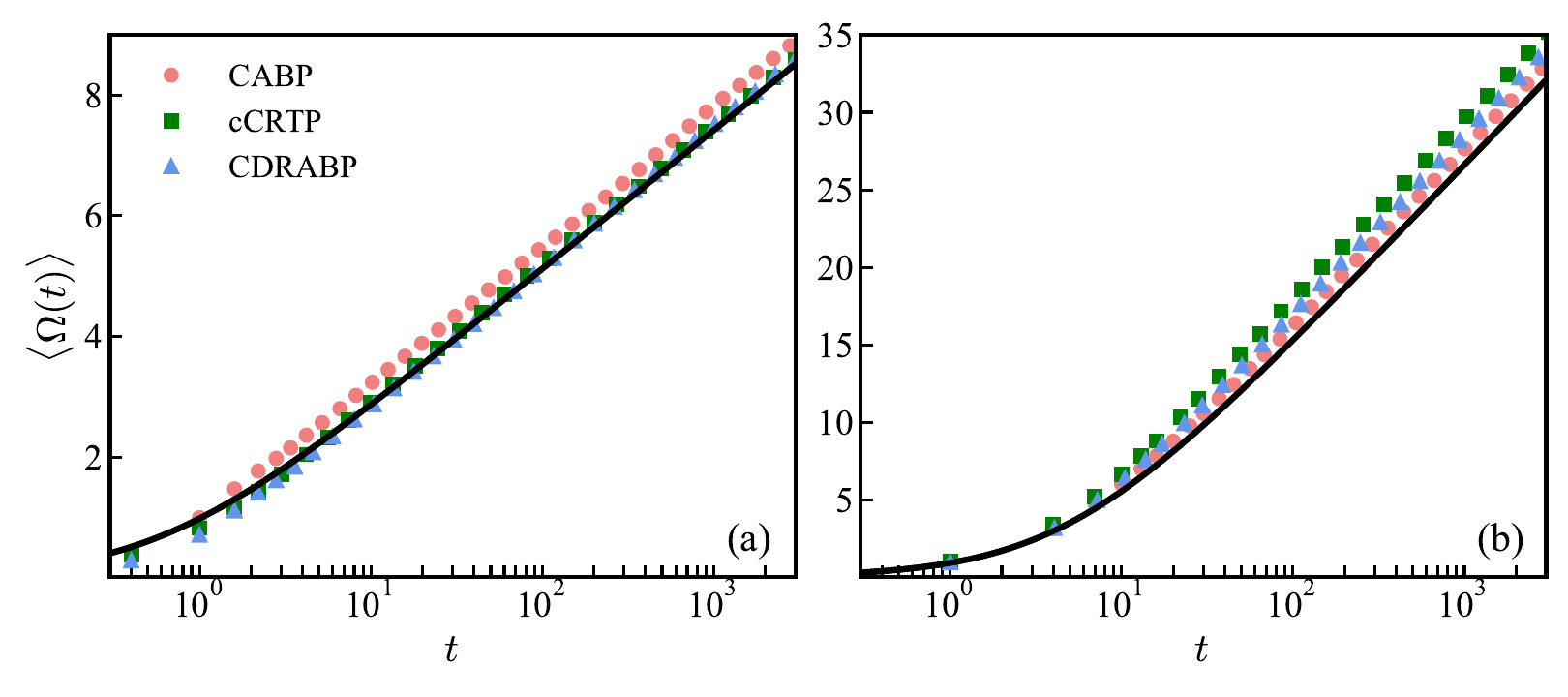}
    \caption{Average of the winding angle, as defined in Eq.~\eqref{eq:def_Om:sm} for the CABP, cCRTP and CDRABP models measured from numerical simulations with (a) $\tau=1$, and (b) $\tau=5$, with  $\omega=2$ and $v_0=1$ for both panels. The black solid line shows the approximate analytical result given by Eq.~\eqref{eq:om-asym:sm}.}\label{fig:omega_comp}
\end{figure}

As expected, the logarithmic growth in Eq.~\eqref{eq:logt} is also seen from the exact numerical evaluation of the integral in Eq.~\eqref{eq:av_Om:sm}. Unfortunately, it is hard to perform the integral over $s$ in Eq.~\eqref{eq:av_Om:sm} explicitly for arbitrary $\omega$ and $\tau$. However, as shown in Fig.~\ref{fig:integrand}(b), for large $\omega\tau$, \eref{eq:int-long} approximates the exact $I(s)$ extremely well, even for small $s <\tau$. Therefore, in the regime of $\omega\tau \gg 1$, we  use \eref{eq:int-long} in Eq.~\eqref{eq:av_Om:sm} and perform the integral explicitly to get
\begin{equation}
 \langle \Omega (t) \rangle \approx  \frac{\omega\tau}{2} \ln (1+ t/\tau_0),  
 \label{eq:om-asym:sm}
\end{equation}
where $\tau_0$ is exactly given by Eq.~\eqref{eq:tau0}. We find that Eq.~\eqref{eq:om-asym:sm} agrees quite well with 
the $\langle \Omega(t) \rangle$ obtained from numerical simulations for large $\omega\tau$, as shown in \fref{fig:omega_comp}.  Evidently, for $t \gg \tau_0$, Eq.~\eqref{eq:om-asym:sm} reduces to Eq.~\eqref{eq:logt}.

For small values of $\omega\tau$ and $r_0$, the parameter $\tau_0$ in Eq.~\eqref{eq:tau0} becomes negative. In this case, as seen from \fref{fig:integrand}~(a), the agreement between Eq.~\eqref{eq:int-long} and the exact $I(s)$ fails for small $s$ due to the divergence of Eq.~\eqref{eq:int-long} at $s=-\tau_0 >0$. Therefore, we cannot use Eq.~\eqref{eq:int-long} to predict the short-time behavior of $\Omega(t)$. Nevertheless, due to the $s^{-1}$ power-law tail of $I(s)$, the long-time behavior in Eq.~\eqref{eq:logt} holds for arbitrary $\omega$ and $\tau$, as shown in the main text.

Figure~\ref{fig:omega_comp} compares $\la \Omega(t) \ra$ obtained from numerical simulations with the prediction \eref{eq:om-asym:sm} for three different models and two different large values of $\omega \tau$. Clearly, \eref{eq:om-asym:sm} describes the large-time behaviour of $\la \Omega(t) \ra$ very well up to a model specific overall  $O(1)$ additive shift. 

\section{Supplementary video}
The typical motion of the chiral mobug with $\omega=2.57$ rad/s, $\tau=0.73$ s, and $v_0 \approx 4.26$ cm/s, is shown in the movie.
The trajectory  $(x(t),y(t))$ of the mobug centroid along with the instantaneous orientation $\theta(t)$, indicated by the arrow, are overlaid on the video recording.